\documentstyle[12pt,aaspp4]{article}
\lefthead{Bureau \& Freeman}
\righthead{The Nature of Boxy/Peanut-Shaped Bulges}
\slugcomment{Accepted for publication in The Astronomical Journal}
\begin{document}
%
%
\newcommand{\coltwo}[2]{\mbox{$\displaystyle\mbox{#1}\atop{\displaystyle\mbox{#2}}$}}
%
%
\title{The Nature of Boxy/Peanut-Shaped Bulges in Spiral Galaxies}
\author{M.\ Bureau\altaffilmark{1} and K.\ C.\ Freeman}
\affil{Research School of Astronomy and Astrophysics, Institute of Advanced
  Studies, The Australian National University, Mount Stromlo and Siding Spring
  Observatories, Private Bag, Weston Creek P.O., ACT~2611, Australia} 
\altaffiltext{1}{Now at Sterrewacht Leiden, Postbus~9513, 2300~RA Leiden, The
Netherlands}
\begin{abstract}
  We present a systematic observational study of the relationship between bars
  and boxy/peanut-shaped (B/PS) bulges. We first review and discuss proposed
  mechanisms for their formation, focussing on accretion and bar-buckling
  scenarios. Using new methods relying on the kinematics of edge-on disks, we
  then look for bars in a large sample of edge-on spiral galaxies with a B/PS
  bulge and in a smaller control sample of edge-on spirals with more
  spheroidal bulges. We present position-velocity diagrams of the ionised gas
  obtained from optical long-slit spectroscopy. We show that almost all B/PS
  bulges are due to a thick bar viewed edge-on, while only a few extreme cases
  may be due to the accretion of external material. This strongly supports the
  bar-buckling mechanism for the formation of B/PS bulges. None of the
  galaxies in the control sample shows evidence for a bar, which suggests
  conversely that bars are generally B/PS.
  
  We consider the effects of dust in the disk of the galaxies, but conclude
  that it does not significantly affect our results. Unusual emission line
  ratios correlating with kinematical structures are observed in many objects,
  and we argue that this is consistent with the presence of strong bars in the
  disk of the galaxies. As expected from $N$-body simulations, the
  boxy--peanut transition appears to be related to the viewing angle, but more
  work is required to derive the precise orientation of the bars the bulges.
  
  The reliable identification of bars in edge-on spiral galaxies opens up for
  the first time the possibility of studying observationally the vertical
  structure of bars.
\end{abstract}
\keywords{galaxies: formation~--- galaxies: evolution~--- galaxies: kinematics and
dynamics~--- galaxies: structure~--- galaxies: spiral~--- instabilities}
\section{Introduction\label{sec:introduction}}
\nopagebreak
Many spiral galaxies display boxy or peanut-shaped bulges when viewed edge-on.
Unfortunately, statistical studies of the incidence of these objects have not
used objective criteria to quantify the boxiness of the bulges, and they are
therefore hard to compare and yield moderately different results.
Nevertheless, it seems clear that at least 20-30\% of edge-on spirals possess
a boxy/peanut-shaped (hereafter B/PS) bulge (see Jarvis \markcite{j86}1986;
Shaw \markcite{s87}1987; de Souza \& dos Anjos \markcite{dd87}1987). Spiral
galaxies with a B/PS bulge are therefore a significant class of objects.

The fact that the boxy/peanut shape is seen only in edge-on spirals indicates
that the shape is related to the vertical distribution of light. Compared to
the usual $R^{1/4}$ light distribution of spheroids, B/PS bulges have excess
light above the plane at large galactocentric radii (see Shaw
\markcite{s93}1993). Furthermore, the three-dimensional (hereafter 3D) light
distribution of B/PS bulges must be even more extreme than their projected
surface brightness (see, e.g., Binney \& Petrou \markcite{bp85}1985 for
axisymmetric models).  B/PS bulges also appear to rotate cylindrically, i.e.\ 
their rotation is independent of the height above the plane (e.g.\ Kormendy \&
Illingworth \markcite{ki82}1982; Rowley \markcite{r86}1986).

Several models have been proposed to explain the structure of B/PS bulges
(e.g.\ Combes \& Sanders \markcite{cs81}1981; May, van Albada, \& Norman
\markcite{mvn85}1985; Binney \& Petrou \markcite{bp85}1985). However, despite
their prevalence and interesting structural and dynamical properties, B/PS
bulges remain poorly studied observationally, probably because the edge-on
projection makes the interpretation of observational data difficult.

In this paper, we present new kinematical data for a large sample of spiral
galaxies with a B/PS bulge. Our main goals are to determine their 3D structure
and to identify their likely formation mechanism.  We use our data to show
that B/PS bulges are simply thick bars viewed edge-on.

In \S~\ref{sec:formation}, we discuss the two main scenarios proposed for the
formation of B/PS bulges -- accretion of satellite galaxies and buckling of a
bar. We describe ways of discriminating between the two scenarios using
kinematical data in \S~\ref{sec:diagnostics}. The observations are presented
in \S~\ref{sec:observations} and the results discussed at length in
\S~\ref{sec:discussion}. We conclude in \S~\ref{sec:conclusions}.
\section{Formation Mechanisms\label{sec:formation}}
\nopagebreak
\subsection{Accretion\label{sec:accretion}}
\nopagebreak
Accretion of external material such as satellite galaxies was the early
favoured mechanism to explain the formation of B/PS bulges in spiral galaxies.
Binney \& Petrou (\markcite{bp85}1985) and Rowley (\markcite{r86}1986,
\markcite{r88}1988) showed that it is possible to construct axisymmetric
cylindrically rotating B/PS bulges from relatively simple distribution
functions. Binney \& Petrou adopted a distribution function with a third
integral of motion, in addition to the energy $E$ and angular momentum along
the symmetry axis $L_z$. This integral favoured orbits reaching a particular
height above the plane. Rowley adopted a two-integral distribution function,
with a truncation depending on both $E$ and $L_z$. Both authors argued that
the required distribution functions can form naturally through the accretion
of material onto a host spiral galaxy. Binney \& Petrou (\markcite{bp85}1985)
additionally argued that, for the accreted material to form a B/PS bulge, the
velocity dispersion of the satellite must be much lower than its orbital
speed, and the decay timescale must be much longer that the orbital time. This
ensures that the accreted material stays clustered in phase space.

Accretion scenarios face several problems. At one extreme, one can consider
the accretion of several small satellite galaxies. However, only a narrow
range of orbital energy and angular momentum can lead to the formation of a
B/PS bulge, so it is improbable that many satellite galaxies would all share
these properties. Remaining satellites should then be present but they are not
observed (Shaw \markcite{s87}1987). The accretion of a single large companion
is also ruled out by the large velocity dispersion and small decay timescale
involved (Binney \& Petrou \markcite{bp85}1985). Similarly, the merger of two
spiral galaxies of similar sizes (or of a spiral and a small elliptical) seems
an unlikely route to form B/PS bulges. This would require a fairly precise
alignment of the spins and orbital angular momenta of the two galaxies. We
recall that about a third of all spiral galaxy bulges should be formed this
way.

From the arguments presented above, it seems that the accretion of a small
number of moderate-sized satellites is the only accretion scenario which may
lead to the formation of B/PS bulges. In favour of accretion is the fact that
the best examples of B/PS bulges are found in small groups (e.g.\ NGC~128,
ESO~597-~G~036). In addition, the possibly related X-shaped galaxies can form
through the accretion of a satellite galaxy (e.g.\ Whitmore \& Bell
\markcite{wb88}1988; Hernquist \& Quinn \markcite{hq89}1989). On the other
hand, no evidence of accretion (arcs, shells, filaments, etc.) were detected
by Shaw (\markcite{s93}1993) near spirals with a B/PS bulge, which argues
against any kind of accretion. Furthermore, spiral galaxies with a B/PS bulge
are not found preferentially in clusters (Shaw \markcite{s87}1987). We are
thus led to the conclusion that, while accretion of external material may play
a role in the formation of some B/PS bulges, it is unlikely to be the primary
formation mechanism.
\subsection{Bar-Buckling\label{sec:buckling}}
\nopagebreak
Bars can form naturally in $N$-body simulations of rotationally supported
stellar disks (e.g.\ Sellwood \markcite{s81}1981; Athanassoula \& Sellwood
\markcite{as86}1986), due to global bisymmetric instabilities (e.g.\ Kalnajs
\markcite{k71}1971, \markcite{k77}1977). Based on 3D $N$-body
simulations, Combes \& Sanders (\markcite{cs81}1981) were the first to suggest
that B/PS bulges may arise from the thickening of bars in the disks of spiral
galaxies. Their results were confirmed and their suggestion supported by later
works (see, e.g., Combes et al.\ \markcite{cdfp90}1990; Raha et al.\ 
\markcite{rsjk91}1991). In short, the simulations show that, soon after a bar
develops, it buckles and settles with an increased thickness and vertical
velocity dispersion, appearing boxy-shaped when seen end-on and peanut-shaped
when seen side-on. The B/PS bulges so formed are cylindrically rotating, as
required.

Toomre (\markcite{t66}1966) first considered the buckling of disks, in highly
idealised models, and found that, if the vertical velocity dispersion in a
disk is less than about a third of the velocity dispersion in the plane, the
disk will be unstable to buckling modes (fire-hose or buckling instability;
see also Fridman \& Polyachenko \markcite{fp84}1984; Araki
\markcite{a85}1985). Bar formation in a disk makes the orbits within the bar
more eccentric without affecting much their perpendicular motions, thereby
providing a natural mechanism for the bar to buckle. Resonances between the
rotation of the bar and the vertical oscillations of the stars can also make
the disk vertically unstable (see Pfenniger \markcite{p84}1984; Combes et al.\ 
\markcite{cdfp90}1990; Pfenniger \& Friedli \markcite{pf91}1991). Irrespective
of exactly how a bar buckled, the final shape of the bar is probably due to
orbits trapped around the 2:2:1 periodic orbit family (see, e.g., Pfenniger \&
Friedli \markcite{pf91}1991).

Bar-buckling is the currently favoured mechanism for the formation of B/PS
bulges. In particular, it provides a natural way to form B/PS bulges in
isolated spiral galaxies, which accretion scenarios are unable to do. A number
of facts also suggest a connection between (thick) bars and B/PS bulges,
although they do not support bar-{\em buckling} directly. In particular, the
fraction of edge-on spirals possessing a B/PS bulge is similar to the fraction
of strongly barred spirals among face-on systems ($\approx30\%$; see, e.g.,
Sellwood \& Wilkinson \markcite{sw93}1993; Shaw \markcite{s87}1987). In a few
cases, a bar can also be directly associated with a B/PS bulge from
morphological arguments. NGC~4442 is such an example (Bettoni \& Galletta
\markcite{bg94}1994).

The main observational problem faced by the bar-buckling mechanism is that
B/PS bulges seem to be shorter (relative to the disk diameter) than real bars
or the strong bars formed in $N$-body simulations. However, this might simply
be due to projection (a given surface brightness level being reached at a
smaller radius in a more face-on disk) and no proper statistics have yet been
compiled. The long term secular evolution of bars is also poorly understood.
For example, it is known that bars can transfer angular momentum to a
spheroidal component very efficiently (e.g.\ Sellwood \markcite{s80}1980;
Weinberg \markcite{w85}1985). On the other hand, Debattista \& Sellwood
(\markcite{ds98}1998) showed that, while a bar can be slowed down, it
continues to grow and the boxy/peanut shape is preserved. It is less clear
what happens if a bar is strongly perturbed. However, Norman, Sellwood, \&
Hasan (\markcite{nsh96}1996) showed that, if a bar is destroyed through the
growth of a central mass concentration (e.g.\ Hasan \& Norman
\markcite{hn90}1990; Friedli \& Benz \markcite{fb93}1993), the boxy/peanut
shape is also destroyed.

The merits of the bar-buckling mechanism significantly outweigh these
potential problems, and the bar-buckling scenario to form B/PS bulges remains
largely unchallenged at the moment, despite very little observational support.
The main aim of this paper is to test this mechanism, by looking for bars in a
large sample of spiral galaxies with a B/PS bulge. Although this stops short
of a direct verification of buckling, it does test directly for a possible
relationship between bars and B/PS bulges. We will come back to this
distinction in \S~\ref{sec:discussion}.

Although it is not part of our program, the Milky Way is a primary example of
such a galaxy. The Galactic bulge is boxy-shaped (Weiland et al.\ 
\markcite{wetal94}1994) and it is now well established that it is bar-like
(e.g.\ Binney et al.\ \markcite{bgsbu91}1991; Paczy\'{n}ski et al.\ 
\markcite{psuskkmk94}1994; see Kuijken \markcite{k96}1996 for a brief review
of the subject).

We note that ``hybrid'' scenarios have also been proposed to explain the
formation of B/PS bulges, and have gathered recent support from statistical
work on the environment of B/PS bulges by L\"{u}tticke \& Dettmar
(\markcite{ld98}1998). An interaction or merger can excite (or accelerate) the
development of a bar in a disk which is stable (or quasi-stable) against bar
formation (e.g.\ Noguchi \markcite{n87}1987; Gerin, Combes, \& Athanassoula
\markcite{gca90}1990; Miwa \& Noguchi \markcite{mn98}1998). Bars formed this
way are then free to evolve to a boxy/peanut shape in the manner described
above (see, e.g., Mihos et al.\ \markcite{mwhdb95}1995), and the bulges thus
formed owe as much to interactions as they do to the bar-buckling instability.
However, the {\em shape} of the bulges is due to the buckling of the bars and
interaction is merely a way to start the bar formation process. In that sense,
hybrid scenarios for the formation of B/PS bulges are really bar-buckling
scenarios, and the possible accretion of material is not directly related to
the issue of the bulges' shape.
\section{Observational Diagnostics\label{sec:diagnostics}}
\nopagebreak
Our main goal with the observations presented in this paper is to look for the
presence of a bar in the disk of spiral galaxies possessing a B/PS bulge.
There is no straightforward photometric way to identify a bar in an edge-on
spiral. The presence of a plateau in the major-axis light profile of the disk
has often been invoked as an indicator of a bar (e.g.\ de Carvalho \& da Costa
\markcite{dd87}1987; Hamabe \& Wakamatsu \markcite{hw89}1989). However,
axisymmetric or quasi-axisymmetric features (e.g.\ a lens) can be mistaken for
a bar and end-on bars may remain undetected. Kuijken \& Merrifield
(\markcite{km95}1995; see also Merrifield \markcite{mk96}1996) were the first
to demonstrate that bars could be identified kinematically in external edge-on
spiral galaxies. They showed that periodic orbits in a barred galaxy model
produce characteristic double-peaked line-of-sight velocity distributions when
viewed edge-on. This gives their modeled spectra a spectacular
``figure-of-eight'' (or X-shaped) appearance, which they were able to observe
in the long-slit spectra of the B/PS systems NGC~5746 and NGC~5965 (see
Fig.~\ref{fig:main} for examples).  Their approach is analogous to that used
in the Galaxy with longitude-velocity diagrams (e.g.\ Peters
\markcite{p75}1975; Mulder \& Liem \markcite{ml86}1986; Binney et al.\ 
\markcite{bgsbu91}1991).

Bureau \& Athanassoula (\markcite{ba99}1999, hereafter BA99) refined the
dynamical theory of Kuijken \& Merrifield (\markcite{km95}1995). They used
periodic orbit families in a barred galaxy model as building blocks to model
the structure and kinematics of real galaxies. They showed that the global
structure of a position-velocity diagram\footnote{Position-velocity diagrams
  (PVDs) show the projected density of material in a system as a function of
  line-of-sight velocity and projected position.} (hereafter PVD) taken along
the major axis of an edge-on system is a reliable bar diagnostic, particularly
the presence of gaps between the signatures of the different families of
periodic orbits. Athanassoula \& Bureau (\markcite{ab99}1999, hereafter AB99)
produced similar bar diagnostics using hydrodynamical simulations of the
gaseous component alone. They showed that, when $x_2$ orbits are present
(corresponding to the existence of an inner Lindblad resonance (hereafter
ILR)), the presence of gaps in a PVD, between the signature of the $x_2$-like
flow and that of the outer parts of the disk, reliably indicates the presence
of a bar. If no $x_2$ orbits are present, one must rely on indirect evidence
to argue for the presence of a bar (see, e.g., Contopoulos \& Grosb\o l
\markcite{cg89}1989 for a review of periodic orbits in barred spirals). The
gaps are a direct consequence of the shocks which develop in relatively strong
bars. These shocks drive an inflow of gas toward the center of the galaxies
and deplete the outer (or entire) bar regions (see, e.g., Athanassoula
\markcite{a92}1992). The simulations of \markcite{ab99}AB99 can be directly
compared with the emission line spectra presented here, and will form the
basis of our argument.

The models of \markcite{ba99}BA99 and \markcite{ab99}AB99 can also be used to
determine the viewing angle with respect to a bar, as the signature present in
the PVDs changes with the orientation of the line-of-sight. In particular,
when a bar is seen end-on, the $x_1$ orbits (and $x_1$-like flow, both
elongated parallel to the bar) reach very high radial velocities, while the
$x_2$ orbits (and $x_2$-like flow, both elongated perpendicular to the bar)
show only relatively low velocities. The opposite is true when a bar is seen
side-on. In addition, the presence or absence of $x_2$ orbits can somewhat
constrain the mass distribution and bar pattern speed of an observed galaxy.

We have not developed specific observational criteria to identify past or
current accretion of material in the observed galaxies. As discussed in
\S~\ref{sec:accretion}, accretion will occur through interactions or merger
events. We will take as the signature of such events, and of possible
accretion, the presence of irregularities in the observed PVDs, in particular
strong asymmetries about the center of an object (see Fig.~\ref{fig:main} for
examples).
\section{Observations\label{sec:observations}}
\nopagebreak
\subsection{The Sample\label{sec:sample}}
\nopagebreak
Our sample of galaxies consists of 30 edge-on spirals selected from the
catalogues of Jarvis (\markcite{j86}1986), Shaw (\markcite{s87}1987), and de
Souza \& dos Anjos (\markcite{dd87}1987) (spiral galaxies with a B/PS bulge),
and from the catalogue of Karachentsev, Karachentseva, \& Parnovsky
(\markcite{kkp93}1993) (spirals with extreme axial ratios; see also
Karachentsev et al.\ \markcite{kkpk93}1993). In order to have enough spatial
resolution in the long-slit spectroscopy, but still be able to image the
galaxies relatively quickly with a small-field near-infrared (hereafter NIR)
camera, we have selected objects with bulges larger than 0\farcm6 in diameter
and disks smaller than about 7\farcm0 (at the 25~B~mag~arcsec$^{-2}$ level).
NIR imaging is important to refine the classification of the bulges and to
subsequently study the vertical structure of the identified bars. All objects
are accessible from the south ($\delta\leq15\arcdeg$). Three-quarters (23/30)
of the galaxies either have probable companions or are part of a group or
cluster. A few of these probably are chance alignments, so it is fair to say
that we are not biased either against or for galaxies in a dense environment.
We should therefore be able to estimate reliably the importance of accretion
in the formation of B/PS bulges.

Of the sample galaxies described above, 80\% (24/30) have a boxy or
peanut-shaped bulge, while 20\% (6/30) have a more spheroidal bulge morphology
and constitute a ``control'' sample. Of the former group, it turned out that
17 galaxies have emission lines extending far enough in the disk to apply the
diagnostics developed by \markcite{ba99}BA99 and \markcite{ab99}AB99 with the
ionised gas; all galaxies in the control group fulfill this condition. In this
paper, we will thus concentrate on a main sample of 17 edge-on spiral galaxies
with a B/PS bulge and a comparison sample of 6 edge-on spiral galaxies with
more spheroidal bulges. The galaxies in each sample are listed in
Tables~\ref{ta:main} and \ref{ta:control} respectively, along with information
on their properties and environment. The galaxies with no or confined emission
are listed in Table~\ref{ta:undetected}. For those, stellar kinematics must be
used to search for the presence of bar. We note that the galaxy type listed in
Tables~\ref{ta:main}--\ref{ta:undetected} is not precise to more than one or
two morphological type, because of the difficulty of classifying edge-on
spirals. The bulge-to-disk ratio is effectively the only criterion left to
classify the objects.

Other than the catalogue of Karachentsev et al.\ (\markcite{kkp93}1993), we
are not aware of any general catalogue of edge-on spiral galaxies. This makes
it difficult to build a large and varied control sample including edge-on
spiral galaxies with large bulges (the Karachentsev et al.\ 
\markcite{kkp93}1993 catalogue is restricted to galaxies with major to minor
axis ratio $a/b\geq7$). Such a catalogue would be very useful, and could
probably be constructed from an initial list of candidates taken from a
catalogue such as the RC3 (de Vaucouleurs et al.\ \markcite{ddcbpf91}1991),
which would then be inspected on survey material.
\subsection{Observations and Data Reduction\label{sec:data}}
\nopagebreak
Our spectroscopic data were acquired between December 1995 and May 1997 (total
of 39~nights) using the Double Beam Spectrograph on the 2.3~m telescope at
Siding Spring Observatory. A $1752\times532$~pixels SITE ST-D06A thinned CCD
was used. The observations discussed in this paper were obtained with the red
arm of the spectrograph centered on the H$\alpha$ $\lambda6563$ emission line.
All galaxies were observed using a $1\farcs8\times400\arcsec$ slit aligned
with the major axis. For objects with a strong dust lane, the slit was
positioned just above it. The spectral resolution is about 1.1~\AA\ FWHM
(0.55~\AA~pixel$^{-1}$) and the spatial scale is 0\farcs9~pixel$^{-1}$. These
data can be directly compared with the gas dynamical models of
\markcite{ab99}AB99.

When no emission line was detected in an object, the red arm of the
spectrograph was moved to the \ion{Ca}{2} absorption line triplet. The blue
arm was always centered on the Mg~$b$ absorption feature. These data will form
the core of a future paper discussing stellar kinematics in the sample
galaxies (including the galaxies in Table~\ref{ta:undetected}). Total exposure
times on both arms ranged from 12000 to 21000~s on each object.

The spectra were reduced using the standard procedure within IRAF. The data
were first bias-subtracted, using both vertical and horizontal overscan
regions, and then using bias frames. If necessary, the data were also
dark-subtracted. The spectra were then flatfielded using flattened continuum
lamp exposures, and wavelength-calibrated using bracketing arc lamp exposures
for each image. The data were then rebinned to a logarithmic wavelength
(linear velocity) scale corresponding to 25~km~s$^{-1}$~pixel$^{-1}$. The
spectra were then corrected for vignetting along the slit using sky exposures.
All exposures of a given object were then registered and offset along the
spatial axis, corrected to a heliocentric rest frame, and combined. The
resulting spectra were then sky-subtracted using source-free regions of the
slit on each side of the objects. The sky subtraction was less than perfect in
some cases, mainly because of difficulties in obtaining a uniform focus of the
spectrograph along the entire length of the slit. This was particularly
troublesome for objects like IC~2531 and NGC~5746 which have sizes comparable
to that of the slit (see, e.g., Fig.~\ref{fig:lineratios}a). In order to
isolate the emission lines, the continuum emission of the objects was then
subtracted using a low-order fit to the data in the spectral direction. The
resulting spectra constitute the basis of our discussion in the next section.

We note that in regions with bright continuum emission, like the center of
some galaxies, the continuum subtraction can leave high shot noise in the
data, which should not be confused with line emission in the grayscale plots
of Figure~\ref{fig:main}--\ref{fig:lineratios}. This is the case for example
in the spectra of ESO~240-~G~011, NGC~1032, and NGC~4703. The effect is
perhaps best seen when a bright star is subtracted, such as in the spectra of
NGC~2788A or NGC~1032 (see Fig.~\ref{fig:main}).

We have not extracted rotation curves from our data. This is because the
entire two-dimensional spectrum, or PVD, is required to identify the signature
of a bar in an edge-on spiral galaxy (see \markcite{ba99}BA99 and
\markcite{ab99}AB99). Evidence of interaction and of possible accretion of
material is also more easily seen in the PVDs. We present the [\ion{N}{2}]
$\lambda6584$ emission line rather than H$\alpha$ because it is not affected
by underlying stellar absorption.
\subsection{Results\label{sec:results}}
\nopagebreak
\placefigure{fig:main}
\placefigure{fig:control}

We present the emission line spectrum for the sample galaxies which have
extended emission only. The PVDs of the galaxies in the main sample and the
control sample are shown in Figure~\ref{fig:main} and
Figure~\ref{fig:control}, respectively. In order to illustrate the range of
galaxy type and bulge morphology in the sample, and to allow connections to be
made between bulge morphologies and kinematical features in the disks, each
PVD is accompanied by a registered image of the corresponding galaxy (from the
Digitized Sky Survey) on the same spatial scale. We discuss here the trends
observed across the data set.

The most important trend, and the main result of this paper, is that most
galaxies in the main sample show a clear bar signature in their PVD (as
described in \S~\ref{sec:diagnostics}). Of the 17 galaxies in the main sample
of spirals with a B/PS bulge and extended emission lines, we conclude that 14
are barred. In these objects, the PVD clearly shows a strong and steep inner
component, associated with an $x_2$-like flow, and a slowly-rising almost
solid-body component, associated with the outer disk, and joining the flat
section of the rotation curve in the outer parts. The two components are
separated by a gap, caused by the absence (or low density) of gaseous material
with $x_1$-like kinematics in the outer bar regions\footnote{The PVDs of
  NGC~128 and IC~2531 do not display a bar signature, but Emsellem \&
  Arsenault (\markcite{ea97}1997) and Bureau \& Freeman (\markcite{bf97}1997)
  showed, using other data, that each galaxy harbours a bar.}. The best
examples of this type of bar signature are seen in the PVD of earlier-type
objects, like NGC~2788A, NGC~5746, and IC~5096. However, the signature is
still clearly visible in the PVD of galaxies as late as ESO~240-~G~011.

In the main sample, only one galaxy, NGC~4469, may be axisymmetric, with no
evidence of either a bar or interaction (although it is not possible to rule
out an interaction which would have occured a long time ago, leaving no
observable trace). Two galaxies, NGC~3390 and ESO~597-~G~036, have a disturbed
strongly asymmetric PVD, which we ascribe to a recent interaction (obvious in
the case of ESO~597-~G~036). These interactions may have led to the accretion
of material. The results for the entire main sample are summarised in
Table~\ref{ta:main}.

Another significant result of this study is that no galaxy in the control
sample shows evidence for a bar. Four of the six galaxies appear to be
axisymmetric, without evidence for either a bar or interaction, and two
galaxies (NGC~5084 and NGC~7123) have a disturbed PVD, indicating they
underwent an interaction recently and possibly accreted material. These
results are tabulated in Table~\ref{ta:control}.
\section{Discussion\label{sec:discussion}}
\nopagebreak
\subsection{The Structure of Boxy/Peanut-Shaped Bulges\label{sec:structure}}
\nopagebreak
The only previous study of this kind was that of Kuijken \& Merrifield
(\markcite{km95}1995), who proposed the method and considered two galaxies;
Bureau \& Freeman (\markcite{bf97}1997) presented preliminary results of the
current work. This is thus the first systematic observational study of the
relationship between bars and B/PS bulges. In summary, our main result is
that, based on new kinematical data, 14 of the 17 galaxies with a B/PS bulge
in our sample are barred, and the remaining 3 galaxies show evidence of
interaction or may be axisymmetric. None of the 6 galaxies without a B/PS
bulge in our sample shows any indication of a bar. This means that most B/PS
bulges are due to the presence of a thick bar viewed edge-on, and only a few
may be due to the accretion of external material. In addition, the more
spheroidal bulges (i.e.\ non-B/PS) do seem axisymmetric. It appears then that
most B/PS bulges are edge-on bars, and that most bars are B/PS when viewed
edge-on. However, the small size of the control sample prevents us from making
a stronger statement about this converse. There is also a continuum of bar
strengths in nature and we would expect to have intermediate cases. The
galaxies NGC~3957 and NGC~4703 may represent such objects: one could argue
that their PVDs display weak bar signatures, and indeed their bulges are the
most flattened in the control sample.

If bars were unrelated to the structure of bulges, we would have expected only
about 5 galaxies in the main sample to be strongly barred, and about 2
galaxies in the control sample (about 30\% of spirals are strongly barred,
Sellwood \& Wilkinson \markcite{sw93}1993). Clearly, our results are
incompatible with these expectations. Recent results by Merrifield \& Kuijken
\markcite{mk99}(1999) also support our conclusions. Based on a smaller sample
of northern edge-on spirals, they show clearly that as bulges become more
B/PS, the complexity and strength of the bar signature in the PVD also
increase.

Our association of bars and B/PS bulges supports the bar-buckling mechanism
for the formation of the latter. However, we do not test directly for
buckling, but rather for the presence of a barred potential in the plane of
the disks. Because bars form rapidly and buckle soon after, on a timescale of
only a few dynamical times (see, e.g., Combes et al.\ \markcite{cdfp90}1990),
it is unlikely that any galaxy in this nearby sample would have been caught in
the act. Thus, other mechanisms which could lead to thick bars cannot be
excluded. In addition, we have no way of determining how the bars themselves
formed, or even whether they formed spontaneously in isolation or through
interaction with nearby galaxies or companions. Therefore, the possibility of
hybrid scenarios for the formation of B/PS bulges, where a bar is formed
because of an interaction and subsequently thickens due to the buckling
instability, remains (see \S~\ref{sec:buckling}).

We have looked mainly at objects with large bulges; only two of the galaxies
studied are late-type spirals (ESO~240-~G~011 and IC~5176). This is a
selection effect due to the difficulty to identify very small B/PS bulges. It
would therefore be interesting to search for bars in very late edge-on spiral
galaxies, and verify whether thin bars do exist: the bar in ESO~240-~G~011 is
not very thick, but it is thicker (isophotally) than the disk. \ion{H}{1}
synthesis imaging is probably the best way to achieve this goal, as these
objects are often dusty and \ion{H}{1}-rich. Such bars may even provide a
novel way to constrain the total (luminous and dark) mass distribution of
spirals, in a manner analogous to the use of warps or flaring, as buckling is
sensitive to the presence of a dark halo (e.g.\ Combes \& Sanders
\markcite{cs81}1981). We will report on \ion{H}{1} synthesis observations of
a few objects in our sample in a later paper.

Our observations revive the old issue of the exact nature of bulges. In
face-on systems, one can often clearly identify a bar and a more nearly
axisymmetric component usually referred to as the bulge. However, Kormendy
(\markcite{k93}1993) has argued that, at least in some examples, these
apparent bulges may just be structures in the disk. In edge-on spirals, we are
not aware of any galaxies displaying two separate vertically extended
components. This raises the question of whether the bars and bulges of
face-on systems are really two distinct structural and dynamical components,
despite the fact that they can be separated photometrically. Our data on
edge-on galaxies tightly link the presence of a bar with the presence of a
B/PS bulge, which suggests that bars and B/PS bulges are very closely related.
This view is supported by theoretical and modelling work on barred spiral
galaxies (e.g.\ Pfenniger \markcite{p84}1984; Pfenniger \& Friedli
\markcite{pf91}1991), as well as by some photometric studies (e.g.\ Ohta
\markcite{o96}1996). However, more work is required to settle the issue.
Kinematical data covering whole bulges would be particularly useful.

\markcite{ba99}BA99 and \markcite{ab99}AB99 proposed diagnostics, again based
on the structure of the observed PVD, to determine the viewing angle with
respect to a bar in an edge-on disk (see \S~\ref{sec:diagnostics}). When the
bar is seen close to side-on, the maximum line-of-sight velocity reached by
the $x_2$-like flow is similar to or larger than the flat portion of the
rotation curve, and the component of the PVD associated with that flow is very
steep. When the bar is seen close to end-on, the $x_2$-like flow only reaches
low velocities and extends to relatively large projected distances. These
diagnostics are ideally suited to test the main prediction of $N$-body models,
that bars are peanut-shaped when seen side-on and boxy-shaped when seen end-on
(see, e.g., Combes \& Sanders \markcite{cs81}1981; Raha et al.\ 
\markcite{rsjk91}1991).

Of the 12 barred galaxies in the main sample for which it is possible to apply
these criteria (we exclude NGC~128 and IC~2531), two-thirds (8/12) seem to be
consistent with the above prediction of $N$-body simulations. For example, in
the galaxy with a peanut-shaped bulge IC~4937, it is clear that the steep
inner component associated with the $x_2$-like flow extends to higher
velocities than the outer parts of the disk (see Fig.~\ref{fig:main}). This
situation is reversed in NGC~1886, which has a boxy-shaped bulge. However,
caution is required when interpreting this result. Firstly, the present
classification of the shape of the bulges is affected by both dust and the low
dynamic range of the material used (Digitized Sky Survey). To remedy this
problem, we have acquired $K$-band images of all the sample galaxies and will
report on these observations in a future paper. Secondly, no clear prediction
has been made from $N$-body simulations about the viewing angle at which the
transition from a boxy to a peanut-shaped bulge occurs. In that regard, it
would be useful to apply quantitative measurements of the boxiness and
``peanutness'' of the bulges to both simulation results and observational data
(see, e.g., Bender \& M\"{o}llenhoff \markcite{bm87}1987 and Athanassoula et
al.\ \markcite{amwpplb90}1990 for two possible methods). Thirdly, because the
$x_2$-like flow is located near the center of the galaxies, the velocities it
reaches depend somewhat on the central concentration of the objects (which
affects the circular velocity in the central regions). This obviously varies
significantly amongst the galaxies in our sample. Therefore, while our
observations may support the prediction of $N$-body models concerning the
orientation of the bar in B/PS bulges, we believe that it is premature to make
a detailed quantitative comparison of the data with the models.

For a more detailed comparison with theory, data of higher signal-to-noise and
higher spatial resolution than the average PVD presented here would be very
desirable. However, it would be worthwhile to model individually the best PVDs
obtained in the present study (e.g.\ NGC~5746, IC~5096, and a few others).
This would very likely lead to tight constraints on the mass distribution and
bar properties of the galaxies, including the orientation and pattern speed of
the bars (see \markcite{ba99}BA99; \markcite{ab99}AB99). The $K$-band images
could also be used to constrain the mass distributions. Comparing the data
with the kinematics (or simply the rotation curve) predicted from an
axisymmetric deprojection would provide an easy test of the shape of the
bulges. On a related subject, the significant thickness of bars suggests that
their 3D structure should be taken into account when deriving the potential of
more face-on systems from NIR images.

In that regard, we should stress that the bar diagnostics we used rely on the
presence of an $x_2$-like flow in the center of the galaxies, and thus on the
existence of ILRs (or, at least, one ILR; see \markcite{ba99}BA99;
\markcite{ab99}AB99). A priori, barred disks or B/PS bulges need not have
ILRs, but at least 13 of the 17 galaxies in the main sample do (we
additionally exclude NGC~128 here). Our data therefore strongly support the
view that barred spiral galaxies generally have ILRs (see also Athanassoula
\markcite{a91}1991, \markcite{a92}1992).
\subsection{Dust and Emission Line Ratios in Boxy/Peanut-Shaped
  Bulges\label{sec:elratios}}
\nopagebreak
Because many galaxies in our sample have a prominent dust lane, it is
important to consider the effects dust may have on our observations. Its
principal consequence in edge-on systems is to limit the depth to which the
line-of-sight penetrates the disk. To bypass this problem, we selected many
galaxies to be slightly inclined, so it was possible in those objects to
position the slit just above the dust lane and have a line-of-sight that still
goes through most of the disk, as required for a comparison with the models of
\markcite{ba99}BA99 and \markcite{ab99}AB99. In the few cases with a strong
dust lane and a perfectly edge-on disk, we tried again to offset the slit
slightly. Unfortunately, with the observational set-up available at the
telescope, it was difficult to position the slit with great precision. The
objects where we suspected that dust could affect our observations are
indicated in Tables~\ref{ta:main}--\ref{ta:undetected}. A large dust optical
depth produces an almost featureless PVD, as one sees only an outer annulus of
material in the disk, and the rotation curve appears slowly-rising and
solid-body (see, e.g., Bosma et al.\ \markcite{bbfa92}1992). The only objects
showing obvious signs of extinction are IC~2531\footnote{This is confirmed by
  the \ion{H}{1} radio synthesis data of Bureau \& Freeman
  (\markcite{bf97}1997), which reveal a complex PVD with a bar signature.},
NGC~4703, and possibly ESO~443-~G~042. Because we see a lot of structure in
most PVDs, including the PVDs of objects with a significant dust lane, we do
not believe that our results are significantly affected by dust. We do detect
a clear bar signature in most galaxies in the main sample.

This statement is strengthened by the fact that all the PVDs showing a bar
signature are close to symmetric. \markcite{ab99}AB99 showed that the bar
signature would be strongly asymmetric in a very dusty disk, and this is not
observed. Similarly, it is improbable that irregular dust distributions would
lead to such well-ordered and symmetric PVDs (very large and localized dust
``patches'' would be required to create the important gaps observed in many
objects).
\placefigure{fig:lineratios}

An unexpected but interesting prospect raised by our observations concerns
emission line ratios. For many of the barred galaxies in the main sample, the
emission line ratios in the central regions are different from those expected
of typical \ion{H}{2} regions. For 9 barred galaxies out of 14, mostly those
with the strongest bar signatures, the [\ion{N}{2}] $\lambda6584$ to H$\alpha$
$\lambda6563$ ratio in the bulge region is greater than unity. In particular,
in a few objects, the steep inner component of the PVD, associated with the
$x_2$-like flow, is much stronger in [\ion{N}{2}] than it is in H$\alpha$,
while the slowly rising component, associated with the outer disk, has a
[\ion{N}{2}]/H$\alpha$ ratio typical of \ion{H}{2} regions. In fact, the inner
component can be almost absent in H$\alpha$. We illustrate these behaviours in
Figure~\ref{fig:lineratios}, which shows the PVDs of the galaxies NGC~5746 and
IC~5096 in the H$\alpha$ and [\ion{N}{2}] $\lambda6548,6584$ lines.

It is possible that the H$\alpha$ emission line is weakened by the underlying
stellar absorption. This suggestion is supported by the fact that the spectra
of many of the galaxies in the main sample show strong Balmer absorption
lines.  However, the H$\alpha$ absorption would need to be very large to
account for the extreme [\ion{N}{2}]/H$\alpha$ ratios observed in some objects
(e.g.\ IC~5096). The strong Balmer lines observed in many objects are
nevertheless interesting in themselves, and indicate that a significant
intermediate age ($\sim1$~Gyr) stellar population is present in the central
regions of the disk of many of the galaxies. It would be interesting to
investigate if these past bursts of star formation can be related to the
presence of the bars.

The high emission line ratios are interesting for two reasons. Firstly, high
[\ion{N}{2}]/H$\alpha$ ratios are commonly believed to be produced by shocks
(see, e.g., Binette, Dopita, \& Tuohy \markcite{bdt85}1985; Dopita \&
Sutherland \markcite{ds96}1996). This is consistent with the view that B/PS
bulges are barred spirals viewed edge-on. The steep inner components of the
PVDs, which display high [\ion{N}{2}]/H$\alpha$ ratios, are associated with an
$x_2$-like flow and the nuclear spirals observed in many barred spiral
galaxies (\markcite{ab99}AB99). Athanassoula (\markcite{a92}1992) showed
convincingly that these are the locus of shocks. Secondly, if one were to
derive H$\alpha$ and [\ion{N}{2}] rotation curves from the data, by taking the
upper envelope of the PVDs (the standard method), the H$\alpha$ and [\ion{N}{2}]
rotation curves would significantly differ for many objects. The [\ion{N}{2}]
lines would yield rapidly rising rotation curves flattening out at small
radii, while the H$\alpha$ line would yield slowly rising rotation curves
flattening out at relatively large radii. Mass models derived from such data
would thus yield qualitatively different results, and our understanding of
galactic dynamics and structure gained from this type of work could be
seriously erroneous (at least for highly inclined spirals). Of course, now
that these galaxies are known to be barred, their rotation curves should not
be used directly for mass modelling, as they are not a good representation of
the circular velocity.

Data such as those presented in Figure~\ref{fig:lineratios} also open up the
possibility of determining the ionisation conditions and abundance of the gas
in different regions of the galaxies in a single observation. Because the
deprojected location of each component of the PVDs is known (see
\markcite{ab99}AB99), this provides an efficient way to study the effects of
bars on the interstellar medium of galaxies on various scales.
\section{Conclusions\label{sec:conclusions}}
\nopagebreak
In this paper, we discussed the various mechanisms proposed for the formation
of the boxy and peanut-shaped bulges observed in some edge-on spiral galaxies.
We argued that accretion scenarios were unlikely to account for most
boxy/peanut-shaped (B/PS) bulges, but that bar-buckling scenarios, discovered
through $N$-body simulations, had this potential. Using recently developed
kinematical bar diagnostics, we searched for bars in a large sample of edge-on
spiral galaxies with a B/PS bulge. Of the 17 galaxies where the diagnostics
could be applied using emission lines, 14 galaxies were shown to be barred, 2
galaxies were significantly disturbed, and 1 galaxy seemed to be axisymmetric.
In a control sample of 6 galaxies with spheroidal bulges, none appeared to be
barred.

Our study supports the bar-buckling mechanism for the formation of B/PS
bulges. Our results imply that most B/PS bulges are due to the presence of a
thick bar that we are viewing edge-on, while only a few may be due to the
accretion of external material. Furthermore, spheroidal bulges do appear to be
axisymmetric. This suggests that all bars are B/PS. Our observations also seem
to support the main prediction of $N$-body simulations, that bars appear
boxy-shaped when seen end-on and peanut-shaped when seen side-on. However,
this issue should be revisited in a more quantitative manner in the future.
With our data, we have no way of determining whether the bars leading to B/PS
bulges have formed in isolation or through interactions and mergers. The
association of B/PS bulges and bars is entirely consistent with the properties
of the bulge of the Milky Way, which is known to be both boxy and bar-like.

We considered the effects of dust on our observations, but concluded that it
does not affect our results significantly. We have also shown that emission
line ratios correlate with kinematical structures in many barred galaxies.
This make possible a direct study of the large scale effects of bars on the
interstellar medium in disks.

Our study opens up the possibility to study observationally the vertical
structure of bars. This was not possible before and represents an interesting
spin-off from the use of bar diagnostics in edge-on spiral galaxies. To this
end, we have obtained $K$-band images of all our sample galaxies, and will
report on this work in a future paper.
\acknowledgments
We thank the staff of Mount Stromlo and Siding Spring Observatories for their
assistance during and after the observations. We also thank A.\ Kalnajs, E.\ 
Athanassoula, A.\ Bosma, and L.\ Sparke for useful discussions. M.\ B.\ 
acknowledges the support of an Australian DEET Overseas Postgraduate Research
Scholarship and a Canadian NSERC Postgraduate Scholarship during the conduct
of this research.  The Digitized Sky Surveys were produced at the Space
Telescope Science Institute under U.S. Government grant NAG W-2166. The images
of these surveys are based on photographic data obtained using the Oschin
Schmidt Telescope on Palomar Mountain and the UK Schmidt Telescope. The plates
were processed into the present compressed digital form with the permission of
these institutions.  The NASA/IPAC Extragalactic Database (NED) is operated by
the Jet Propulsion Laboratory, California Institute of Technology, under
contract with the National Aeronautics and Space Administration.
\clearpage
%
%
%
\figcaption{Structure and kinematics of the galaxies in the main sample of
  spirals with a boxy/peanut-shaped bulge. Each plot corresponds to a galaxy.
  For each, the top panel shows a blue image of the galaxy (identified in the
  bottom-left corner) from the Digitized Sky Survey, and the bottom panel
  shows an ionised gas ([\ion{N}{2}] $\lambda6584$ emission line)
  position-velocity diagram (PVD) taken along the major axis, and
  registered with the image above it. The galaxies are ordered as in
  Table~\ref{ta:main}.
\label{fig:main}}
%
%
\figcaption{Same as Figure~\ref{fig:main} but for the control sample of
  galaxies (Table~\ref{ta:control}).
\label{fig:control}}
%
%
\figcaption{Structure and kinematics of (a) NGC~5746 and (b) IC~5096. For each
  galaxy, the top panel shows a blue image of the galaxy from the Digitized
  Sky Survey, and the bottom panel shows ionised gas position-velocity
  diagrams (PVDs) taken along the major axis, and registered with the image
  above them. The PVDs correspond to the [\ion{N}{2}] $\lambda6584$, H$\alpha$
  $\lambda6563$, and [\ion{N}{2}] $\lambda6548$ emission lines (from top to
  bottom). Note the large [\ion{N}{2}] $\lambda6584$ to H$\alpha$ ratio in the
  inner steep component of the PVDs.
\label{fig:lineratios}}
%
%
%
%
%
\begin{deluxetable}{llllll}
\tablewidth{0pt}
\tablecaption{Main Sample (Detections)\label{ta:main}}
\tablehead{\colhead{Galaxy} & \colhead{Type} & \colhead{Bulge} &
\colhead{Environment} & \colhead{Structure} & \colhead{Notes}}
\startdata
NGC~128 & S0 pec & Peanut & In group & Bar\tablenotemark{a} & \nodata \nl
ESO~151-~G~004 & S0$^0$ & Peanut & In group & Bar & \nodata \nl
NGC~1886 & Sab & Boxy & Isolated & Bar & \nodata \nl
NGC~2788A & Sb & Peanut & In cluster & Bar & Dusty \nl
IC~2531 & Sb & Peanut & In cluster & Bar\tablenotemark{b} & Dusty \nl
NGC~3390 & Sb & Boxy & Companions? & Accretion & \nodata \nl
NGC~4469 & SB(s)0/a? & Peanut & In cluster & Axisymmetric & \nodata \nl
NGC~4710 & SA(r)0$^+$ & Boxy & In cluster & Bar & \nodata \nl
PGC~44931 & Sbc & Peanut & Isolated & Bar & \nodata \nl
ESO~443-~G~042 & Sb & Peanut & Companions? & Bar & Dusty \nl
NGC~5746 & SAB(rs)b? & Peanut & In group & Bar & \nodata \nl
NGC~6722 & Sb & Peanut & Isolated & Bar & \nodata \nl
NGC~6771 & SB(r)0$^+$? & Peanut & In group & Bar & \nodata \nl
IC~4937 & Sb & Peanut & In cluster & Bar & Dusty \nl
ESO~597-~G~036 & S0$^0$ pec & Peanut & In group & Accretion & Dusty \nl
IC~5096 & Sb & Boxy & Companions? & Bar & \nodata \nl
ESO~240-~G~011 & Sb & Boxy & In group & Bar & \nodata \nl
\enddata
\tablenotetext{a}{Emsellem \& Arsenault (\markcite{ea97}1997)}
\tablenotetext{b}{Bureau \& Freeman (\markcite{bf97}1997)}
\end{deluxetable}
\clearpage
%
%
%
\begin{deluxetable}{llllll}
\tablewidth{0pt}
\tablecaption{Control Sample\label{ta:control}}
\tablehead{\colhead{Galaxy} & \colhead{Type} & \colhead{Bulge} &
\colhead{Environment} & \colhead{Structure} & \colhead{Notes}}
\startdata
NGC~1032 & S0/a & Spheroidal & Companions? & Axisymmetric & Dusty \nl
NGC~3957 & SA0$^+$ & Spheroidal & In cluster & Axisymmetric & \nodata \nl
NGC~4703 & Sb & Spheroidal & Isolated & Axisymmetric & Dusty \nl
NGC~5084 & S0 & Spheroidal & In cluster & Accretion & \nodata  \nl
NGC~7123 & Sa & Spheroidal & Isolated & Accretion & Dusty \nl
IC~5176 & SAB(s)bc? & Spheroidal & Companions? & Axisymmetric & \nodata \nl
\enddata
\end{deluxetable}
\clearpage
%
%
%
\begin{deluxetable}{llllll}
\tablewidth{0pt}
\tablecaption{Main Sample (Non-Detections)\label{ta:undetected}}
\tablehead{\colhead{Galaxy} & \colhead{Type} & \colhead{Bulge} &
\colhead{Environment} & \colhead{Notes}}
\startdata
NGC~1381 & SA0 & Boxy & In cluster & \nodata \nl
NGC~1596 & SA0 & Boxy & In group & \nodata  \nl
NGC~2310 & S0 & Boxy & Isolated & \nodata \nl
ESO~311-~G~012 & S0/a? & Boxy & Isolated & \nodata \nl
NGC~3203 & SA(r)0$^+$? & Boxy & In group & \nodata \nl 
IC~4767 & S pec & Peanut & In cluster & \nodata \nl
ESO~185-~G~053 & SB pec & Peanut & In cluster & \nodata \nl
\enddata
\end{deluxetable}
%
%
%
%
%
%
%
%
%
%
%
%
%
%
%
%
\end{document}